\documentclass[preprint,amsmath,amssymb,aps]{revtex4}

\usepackage{graphicx}
\usepackage{subfigure}
\newcommand{\notes}[1]{}

\begin{document}

\title{Mapping the train model for earthquakes onto the stochastic sandpile model}

\author{C. V. Chianca} \author{J. S. S\'a Martins} \author{P. M. C. de Oliveira} 
\affiliation{Instituto de F\'\i sica, Universidade Federal Fluminense, 24210-346 Niter\'oi-RJ, Brazil \\ and National Institute of Science and Technology for Complex Systems}

\date{\today}

\begin{abstract}

We perform a computational study of a variant of the ``train'' model for earthquakes [{\it \pra} {\bf 46}, 6288 (1992)], where we assume a static friction that is a stochastic function of position rather than being velocity dependent. The model consists of an array of blocks  coupled by springs, with the forces between neighbouring blocks balanced by static friction. We calculate the probability, $P(s)$, of the occurrence of avalanches with a size $s$ or greater, finding that our results are consistent with the phenomenology and also with previous models which exhibit a power law over a wide range. We show that the train model may be mapped onto a  stochastic sandpile model and study a variant of the latter for non-spherical grains. We show that, in this case, the model has critical behaviour only for grains with large aspect ratio, as was already shown in experiments with real ricepiles. We also demonstrate a way to introduce randomness in a physically motivated manner into the model.

\end{abstract}

\maketitle

\bigskip

\section{Introduction}
\label{sec:introduction}

Earthquakes are the subject of intensive study. The interest in the study of these phenomena can be either academic, triggering the development of models which help to unveil their basic characteristics, or highly practical, with the objective of minimising the effects of massive destruction caused by extreme events. In line with practical thinking, modelling can be of importance when one tries to understand the dynamical causes of earthquakes, with the eventual goal being that of forecasting. The successful achievement of this goal would have obvious economic consequences, especially for some of the heavily populated regions which are subject to seismic activity. One difficulty which is present in the study of earthquakes is the impossibility of direct observational access to the microscopic dynamics of tectonic faults, leaving us without a complete physical description of their occurrence. The only data that can be obtained are the result of measurements of crust
displacements and of seismic waves, rather than of the close-up dynamics of the fault movements.

Any model or theory that intends to explain the occurrence of earthquakes should also give a satisfactory explanation of the large quantity of statistics which have been gathered. It is well known that real earthquakes follow empirical power laws for the frequency distribution of their magnitudes and for the temporal decay of aftershock rates. These laws are known as the Gutenberg-Richter scaling law and Omori's law~\cite{GR,Omori}. 

Various models based on the observed evidence for large scale motions of the Earth's crust, that is, on the dynamics of tectonic plates sliding slowly against each other, have been proposed to investigate earthquakes~\cite{BK,OFC,Carlson,Souza-Vieira,Rundle}. The treatment of these problems from the point of view of friction leads to a nonlinear description of the phenomena. To describe the mechanisms of continuous media we generally would solve a set of partial differential equations, but these solutions can become extremely expensive in terms of computational time. It is usual to simplify the problem by picturing locations on opposite sides of a fault as a two-dimensional network of masses connected by springs, which model the elastic interactions, and pinned down by static friction. Once we have discretised the problem, we are also free to use cellular automata techniques, which allow us to work with larger systems at less computational cost.

In 1967, Burridge and Knopoff~\cite{BK} proposed a model based on the picture described above, which imitates the ``stick-slip'' dynamics of real events~\cite{Rubinstein,Chelidze}, and whose dynamics could be solved numerically, referred to in this manuscript as the BK model. The system they studied consisted of a chain of masses coupled by springs and in contact with a rough surface. Each connecting spring represents a continuous section of a fault line, giving a linear elastic coupling between the blocks that represent the irregularities on the fault interface. This was the precursor of the train model, where the spring connected to the first block generates the instability. The results for the BK and similar models gave a behaviour for the distribution of events which agreed with the Gutenberg-Richter Law. The BK models and others developed subsequently use a velocity dependent friction force to produce a dynamical instability, which then leads to complex dynamical behaviours~\cite{Ferguson,Pradhan}. Another model which is phenomenologically similar to real seismic faults was proposed by Olami et al~\cite{OFC}, and is known as the OFC model. This model uses a dynamical field, which is thought to represent local forcing, over a regular lattice. The value of the field is updated synchronously over the whole lattice at discrete time intervals, rising monotonically and uniformly up to a threshold value. Once this threshold is reached at one site, the site is said to become unstable and the value of the field at this site is reset to some residual value, often taken as zero, while a fraction of the value by which it is decreased is distributed among its neighbours. If this fraction is smaller than unity, the model is said to be non-conservative. Complex behaviour can arise due to clustering and synchronisation of the field variable, leading to a cascade of sites becoming unstable in sequence. The choice of neighbours of a site can either be performed once, obeying the regular metric of the lattice, or randomly, a new list of neighbours being chosen at each updating. The latter method of redistributing neighbours destroys any spatial correlations~\cite{Jensen}. Even when this model uses only short range local interactions, long range correlations appear due to criticality, and it still exhibits a behaviour similar to the empirical laws. We note that models with long range interactions have also been analysed and show behaviours which follow the empirical phenomenology~\cite{Rundle}. In general, when one treats sliding blocks, the overdamped regime is considered, although the underdamped regime has also been studied~\cite{Elmer}. By operating in the overdamped regime, inertia is neglected, so that it plays no role in the dynamics. By so doing, any wave mechanisms for the relaxation of energy and stress are not taken into account. Because the energy carried by wave movements does not exceed 10$\%$ of the total liberated in an event, this approximation is usually considered to be acceptable.

In line with the ideas discussed above, this paper develops and uses a spring-block model where the equilibrium position of each block on the surface depends on the balance between the local stochastic maximum static friction and the elastic forces. We find numerical evidence that our model has similar descriptive powers to the Burridge-Knopoff model. We also show how our model can be mapped onto a sandpile model which has the Oslo model~\cite{Christensen} as limiting case, reinforcing the conjecture that both these models belong to the same universality class.

\section{Description of a train model variant}

The train model, introduced in 1992 by Souza-Vieira~\cite{Souza-Vieira}, is a mechanical model with blocks and springs, inspired by the Burridge-Knopoff experiment, for dynamic earthquake simulation. The model consists of a one-dimensional chain of blocks connected by springs. Each block is in contact with a lower rough surface, and the chain is pulled at one end with a small constant velocity. This model is completely deterministic. In order to study the effect of  velocity independent friction we develop here a discrete stochastic version of the train model. A model with stochastic friction such as ours has previously been proposed to study the  characteristic dynamics of a block sliding on a rough inclined plane~\cite{Lima}. We will focus on the displacement between points where the elastic forces are balanced by static friction between the blocks and the surface. The blocks are joined by ideal and equal springs. The stability is established solely by the balance of the three forces acting on each block - exerted by the two springs and the local static friction. Thus, whenever $(\sum \vec{F})_{i} > 0$, block $i$ moves until the next point where $(\sum \vec{F})_{i} = 0$, with $(\sum \vec{F})_{i} < 0$ never occurring. Fundamentally, an earthquake is a fracture that does not proceed instantaneously, but is initiated locally and propagates rapidly across the surface of the fault. It eventually stops, either due to energy dissipation or when it encounters a  stronger asperity, or point of contact. We model these asperities as random points of contact between the two surfaces. The maximum static friction force, $M_{i}$, is due to these randomly scattered contact points between the block and the surface. We represent the roughness of both surfaces by means of binary strings of $0$s and $1$s. Each bit can be thought of as representing the average properties of the surface over an arbitrarily small length. If, for instance, a certain region is more prominent than the average, the corresponding bit is set to $1$, and to $0$ in the opposite case~\cite{Lima}. Thus, when the block is put in contact with the surface, the only regions which contribute to the frictional forces are those in which both have $1$ at the corresponding locations of their bit strings, and $M_{i}$ is the number of such coincidences.

\begin{figure}
\centering
\includegraphics[width=14.cm]{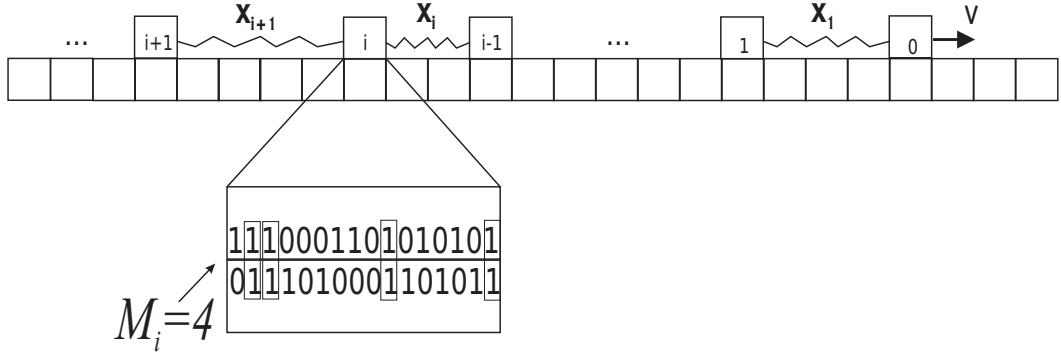}
\caption{Schematics of the ``train model''. The expanded section shows how the friction is calculated.}
\label{fig:Figure1-1}
\end{figure}

The dynamics of the system are as follows: a chain of blocks is placed over a surface in some initial configuration. The rightmost block, or block $0$, nicknamed the ``engine'', moves at a constant speed to the right, and pulls its left neighbour. The engine's motion is the generator of instability in the system. The simulation starts with global equilibrium: the total force on each block is $0$. At each step, the engine moves one unit distance to the right and a time counter is incremented. After some steps, its left neighbour is driven into an unstable state and an avalanche event sets in. For speeding up the actual simulation, the engine is moved right the actual number of steps needed to unbalance its left neighbour, and the time counter updated accordingly. During the avalanche, the positions of the blocks are updated as follows: Block $i$ sits over some entry of this chain.  Block $i-1$ is $X_{i}$ positions to the right and block $i+1$ is $X_{i+1}$ positions to the left of block $i$, pulling block $i$, which will move a number of unit steps to the right until $\Delta X_{i} = X_{i} - X_{i+1} \leq M_i$, where $M_i$ is the position-dependent maximum friction between block $i$ and the surface currently below it. Sucessive blocks are updated from right to left, scanning the whole chain as many times as needed to reach a global equilibrium. After that, an avalanche is complete, and the engine is again moved one position to the right and the above dynamics is repeated for the next avalanche.  We illustrate this model in Fig.~\ref{fig:Figure1-1}. For the purpose of establishing the cumulative probability distribution of the model, we may define the size of an avalanche in two different ways; either as being the number of block movements within an avalanche or as the total number of blocks which move during the whole event - the same block may move more than once. Since our results show the same qualitative behaviour in either case, we will stick to the first definition in what follows. We consider that an avalanche is not over until all the blocks are again in a stable configuration.

\begin{figure}
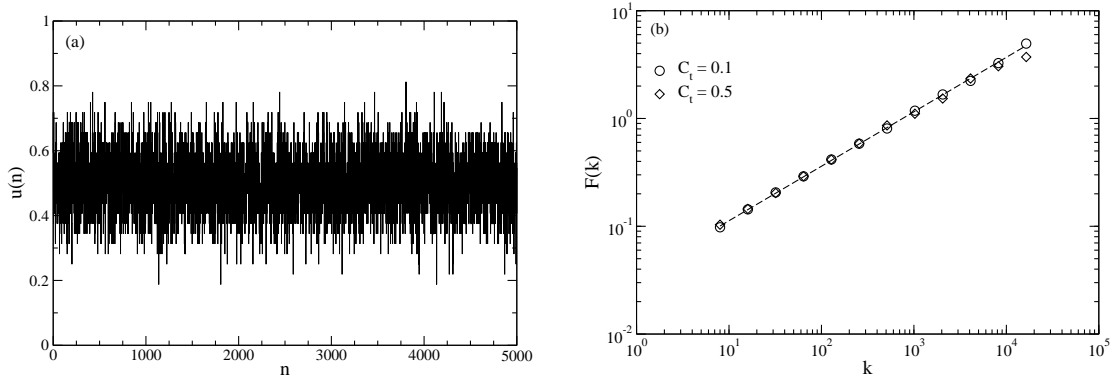

\centering
\subfigure{\includegraphics[width=7.cm]{Serie05.eps}
\label{fig:serie05}}
\hspace{8pt}
\subfigure{\includegraphics[width=7.cm]{Expoente.eps}
\label{fig:expoente}}
\caption{\subref{fig:serie05} The first $5000$ terms of the series $u(n)$ of superposition of the surface chains, where $n=1,\dots, N_{max}$ and $N_{max}$ is the length of the series. Also $u(n)$ is not an integer because it is divided by the total number of bits in each chain; \subref{fig:expoente} The fluctuation function $F(k)$ for the superposition function of the surface chains. The exponents give approximately $ 0.5$, characteristic of white noise.}
\label{fig:Correl}
\end{figure}

Each block and each surface position is represented by a random bit-string, fixed during all the simulation. We are free to vary the size of these strings, allowing them to be $1,2,4,\dots,32$ or $64$ bits long. The surface is formed when we collect the strings alongside each other, with typical total lengths being of the order of $10^6$ bits. In order to construct the bit-strings along the surface, first a common random pattern is created, the same for all surface positions. Each bit along the surface is then flipped with probability $C_{t}$. Another and independent random pattern is also created for the blocks. Each bit is then flipped with probability $C_{b}$. The amount and positions of bits set to $1$ in these bit chains represent both the number and positions of the asperities in the system. The formation of these patterns for the bit chains representing the blocks and the surface does not introduce correlations into the system. This can be seen by an analysis of the series $u(i)$ resulting from the overlap between the chain of blocks and the surface. An example is shown in Fig.~\ref{fig:serie05} with $C_{t} = 0.5$. This superposition gives us the number of bits that coincide between the different chains. To know whether there is any correlation in the series we calculate the fluctuation function, $F(k)$, and find its correlation exponent, where $k$ is the box size used to calculate the fluctuation function. We find that these exponents are practically identical to the random value $1/2$ for both $C_{t} = 0.1$ and $C_{t} = 0.5$, showing that there is no correlation [see Fig.~\ref{fig:expoente}].

We collect data which allow us to calculate the cumulative probability distribution, $P(s)$, of events that are larger than a size $s$, as a function of this size. Another set of simulations were run in which the concentration of bits in the chains was varied; in other words, we varied $C_t$ and $C_b$ and checked that both the qualitative and quantitative characteristics of the resulting distributions were not modified by the introduction of more bits set to $1$s into the system during the initialisation process.

\begin{figure}
\centering
\includegraphics[width=8.cm]{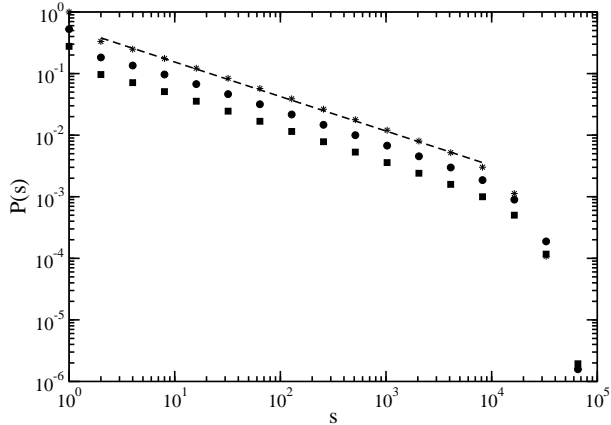}
\caption{Log-log plot of the cumulative probability distribution $P(s)$ of finding avalanches larger than $s$, as a function of $s$ in our variant of the train model. Averages are taken from $6 \times 10^{6}$ events.  Each of the three curves is an average of $30$ runs and  represents different bit concentrations, with ($\ast$) for $C_{t} = 0.1$ and $C_{b} = 0.5$, ($\blacksquare$) for $C_{t} = 0.3$ and $C_{b} = 0.2$, and $(\bullet)$ $C_{t} = 0.5$ and $C_{b} = 0.5$. The two last curves were displaced vertically for the purpose of better visualization. The dotted line corresponds to a power-law with exponent $0.55$. }
\label{fig:Fig1}
\end{figure}

In Fig.~\ref{fig:Fig1} we show the results of sampling $6 \times 10^6$ events, after waiting for a transient of $6 \times 10^6$ avalanches. We find a Gutenberg-Richter power law behaviour, and evaluate the power law exponent as $0.55 \pm 0.04$ for the cumulative probability distribution. This was obtained through data collapse and averaged over exponentially increasing bins with base 2. Despite the simplicity of this model, it compares favourably with previous models, such as the 1-dimensional boundary driven Oslo model \cite{Christensen}, as well as with data collected from real events \cite{GR}.  Although our model uses a stochastic friction force, differently from the deterministic train models, we find a value of the exponent consistent with those found in other models. Our results suggest that the velocity decreasing friction force is not a determining factor for this dynamically complex behaviour, as has been suggested in previous models~\cite{Souza-Vieira,Carlson}.

If we allow violation of Newton's third law, the resulting model enables us to run simulations of much larger systems, for which we also obtain a power law behaviour for the probability distributions of the size of the avalanches. We allow for this violation by supposing that each block is only pulled from one side of the chain. After the  rightmost block is pulled, we begin to check the balance in each block. In this case the comparison is made between the block and its immediately previous neighbour. If this force is smaller than the (maximum) friction on the block, it will be in balance. This makes the simulation more advantageous computationally because we need to sweep the whole chain only once to achieve complete balance of the system after the triggering of each avalanche. The power law exponents for this variant are not the same as for the previous one: for the cumulative probability distribution, we found it to be $0.33$ (figure not shown). We will stick to the first variant in what follows.

In our variants of the train model, the way in which the bit-strings are formed does not allow us to manipulate directly the distribution of the white noise (static friction), and its variance in particular. The question of whether this model is sensitive to the variance of the noise, a phenomenon that shows up rather vividly in other instances, such as in noise-induced ordering~\cite{Anderson}, may be addressed only indirectly in this context. To circumvent this inconvenience one could abandon the idea of generating the noise dynamically, as done so far, and extract it from a predefined distribution with any chosen variance. In our case, this strategy would lead us to pick $M_{i}$ anew from that distribution every time a block moves . Unfortunately this algorithm weakens the appeal of the model we have used so far. Noting that, with noise being generated directly from a distribution, there is a perfect mapping onto a sandpile model, we prefer to examine the impact of the noise distribution using the language of sandpiles, with the extra benefit that it will facilitate a comparison between our results and what is already known about this class of models in the literature.

\section{Mapping onto a sandpile model}
\label{sec:}

We will now map our model onto a sandpile model~\cite{Christensen,Frette,Paczuski,Dickman}, as described in the following. We consider a one-dimensional sandpile of width $L$ and an integer variable, $h_{i}$, which represents the number of grains at each site, $i=1,\dots,L$. The number of blocks in the train model is mapped onto the number of sites of the sandpile, whose first site (site $0$) represents the engine's neighbour. The distance between block $i$ and its right neighbour $X_{i}$ is mapped onto the height of the pile at site $i$, $h_{i}$, leading the local slope of the pile $z_{i} = h_{i} - h_{i+1}$ to represent the net elastic force on block $i$ of the train model, $z_{i} = X_{i} - X_{i+1}$. A site $i$ becomes unstable when its local slope $z_{i}$, which is now the basis for stability criteria, exceeds a site-dependent threshold value $z_{i}^{c}$. In this event, a grain is moved from the unstable site to its downstream neighbour, decreasing the local slope by $2$ units by increasing $X_{i+1}$ and decreasing $X_{i}$ both by $1$ unit. The latter may in turn become unstable and start a cascade of instability along the pile. The threshold slope $z_{i}^{c}$, whose role in the sandpile model is equivalent to that of the maximum static friction force $M_{i}$ in the train model, is reset each time a grain tumbles from site $i$ onto its neighbour and is chosen randomly from some distribution. The sandpile dynamics are driven by dropping a grain at a time on the pile's first site. Again, for speeding up the simulation run, at the end of each cascade the exact number of grains needed to unbalance the engine's neighbour $n_{add} = z_{0}^{c} - z_{0} + 1$ is added to the pile's first site,

\begin{equation}
  h_{0} \rightarrow h_{0} + n_{add}  \quad \text{leading to} \quad z_{0} \rightarrow z_{0}^{c} + 1.
\end{equation}

In this manner, each run of the model is equivalent to a block being pulled with a constant velocity. As a site $i$ becomes unstable one grain tumble from it to the site to the right of it,

\begin{equation}
 h_{i} \rightarrow h_{i} - 1 \text{, } h_{i+1} \rightarrow h_{i+1} + 1 \quad \text{leading to} \quad z_{i} \rightarrow z_{i} - 2 \text{, } z_{i \pm 1} \rightarrow z_{i \pm 1} + 1
\end{equation}

\begin{figure}[]
\centering
\includegraphics[width=12.cm]{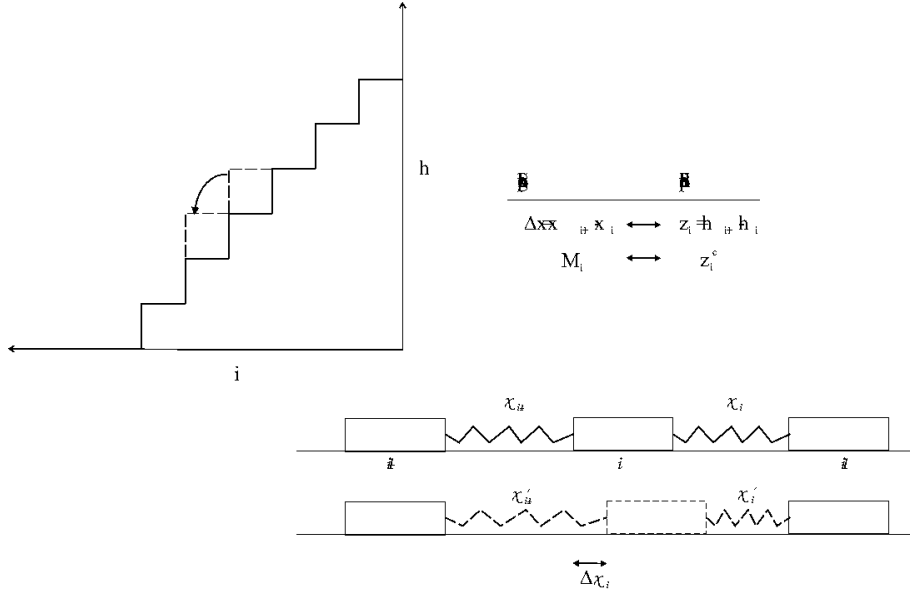}
\caption{Schematics showing the mapping between the spring-block and sandpile models.}
\label{fig:Figure1-2}
\end{figure}

Since the last ``car'' has no left neighbour, the train model is equivalent to an open sandpile, $h_{L+1} = 0$. No grains are added to the pile during an avalanche, which stops when the system reaches a globally stable state with $z_{i} < z_{i}^{c}$ for each $i$.  Grains are then added and a new avalanche is initiated. In Fig.~\ref{fig:Figure1-2}, we give a schematic picture of the sandpile and show the correspondence with the variables of the train model.

\begin{figure}
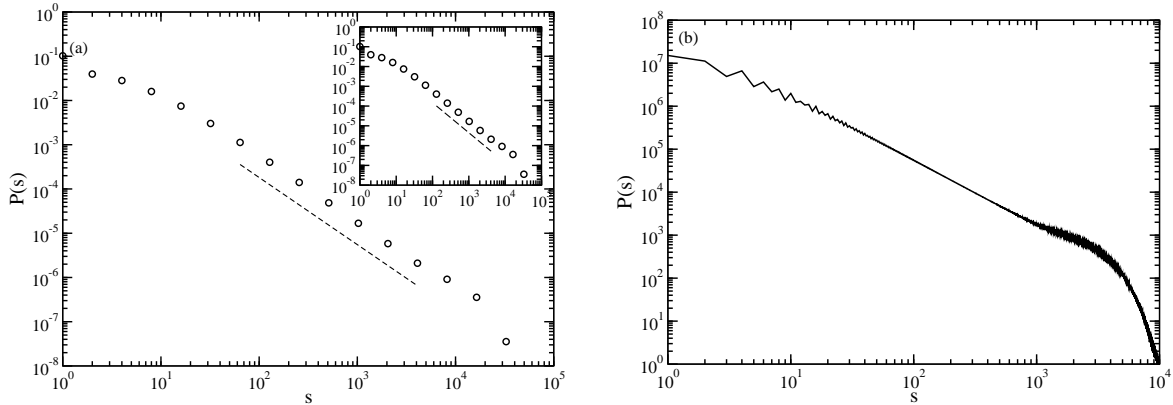

\centering
\subfigure{\includegraphics[width=0.45\textwidth]{Figure5A.eps}
\label{fig:Fig5}}
\hspace{8pt}
\subfigure{\includegraphics[width=0.45\textwidth]{Figure6A.eps}
\label{fig:Fig4}}
\caption {\subref{fig:Fig5} Non-cumulative distributions of avalanche sizes for the train model with blocks with 32 bits each. Inset: Result for sandpile model using the same values for corresponding variables. Both show power law behaviour over three decades, with an exponent of $1.52$. 
\subref{fig:Fig4}  Non-cumulative distribution of avalanche sizes for the Oslo model $z_{i}^{c} \in [1,2]$.}
\label{fig:General}
\end{figure}

We now compare results obtained in simulations of these two dynamics. For the open sandpile we chose for $z_{i}^{c}$, an integer random variable, a uniform distribution in $[1,16]$, which should be compared with the train model with 32 bits for each block. Statistics for the cumulative frequency distribution of the avalanche sizes were collected and are shown in Fig.~\ref{fig:Fig5}. The results are equivalent, as expected. When the uniform distribution for $z_{i}^{c}$ is taken in the integer interval $[1,2]$ the sandpile is known as the Oslo model~\cite{Frette}. The frequency distribution for this limit is presented in Fig.~\ref{fig:Fig4} and shows a saw-tooth behavior for small sizes of the avalanches.

The question about the critical behavior of real sandpiles in nature was addressed in a seminal work describing experiments made with piles of rice grains~\cite{Frettea}. There it was shown that the frequency distribution of avalanches may or may not show the signature of critical behavior when grains of different aspect ratios (length/width) were used. The introduction of the aspect ratio in the model allows us to simulate situations where the grains are no longer isotropic. The anisotropy of the grains generates a variety of packing situations \cite{Frettea}. We can therefore simulate systems with different aspect ratios and examine the behavior of the frequency distribution of avalanches that result. We establish the width of the grain as our unit, and its length is then given by the aspect ratio. In this case, the dynamics has to be modified. We still feed the system by driving its first site to instability, allow for its relaxation, and then check for further instabilities in the slope generated by grain redistribution. If the slope at some site $i$ is larger than $z_{i}^{c}$, $z_{i} >  z_{i}^{c}$, one grain is moved to the right along the sandpile, $h_{i} \rightarrow h_{i} - [1,q]$, $h_{i+1} \rightarrow h_{i+1} + [1,q]$, where $[1,q]$ is a random number between the width and the length of the grain, since its width is our unit of length. This strategy simulates a random orientation of the grain, both in the site where it came from and in that to which it moved. This relaxation rule is used until all sites have a stable slope, $z_{i} \leq z_{i}^{c}$, before a new grain is added to the first site.

\begin{figure}
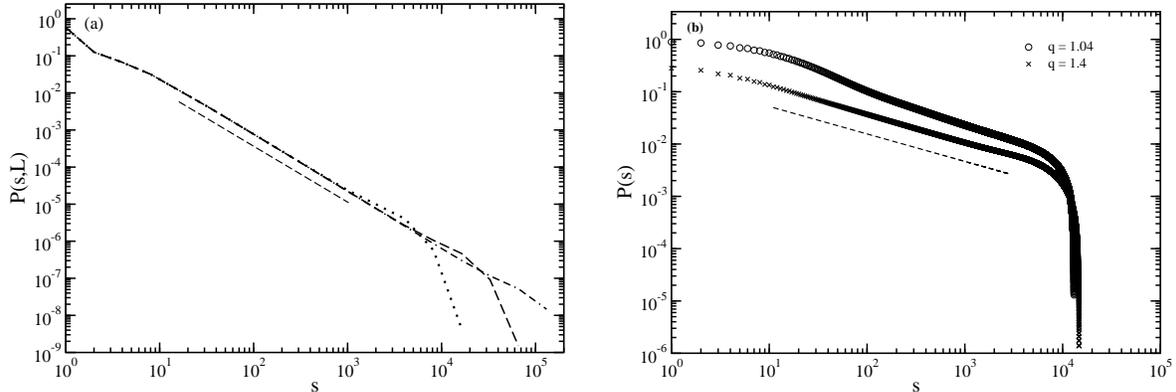

\centering
\subfigure{\includegraphics[width=0.45\textwidth]{Figure6.eps}
\label{fig:Fig6}}
\hspace{8pt}
\subfigure{\includegraphics[width=0.45\textwidth]{Figure7.eps}
\label{fig:Fig7}}
\caption {\subref{fig:Fig6} Non-cumulative distributions of avalanche sizes for the Oslo model with real variables for lattice sizes $L = 100, 200, 400$ and for spherical grains ($q=1.0$). The straight line is a power law with exponent $1.52$.
\subref{fig:Fig7} The log-log cumulative distribution of avalanche sizes in the sandpile model with real variables and elongated grains. The critical and noncritical behaviour is shown for different aspect ratios. The plot for the larger aspect ratio was displaced vertically downward for better visualization. The straight line is a power law with exponent $0.52$.}
\label{fig:Oslo}
\end{figure}

In Fig.~\ref{fig:Fig6} we show results for these dynamics with systems of different sizes and a real-valued $z_{i}$. The simulations with different aspect ratios is shown in Fig.~\ref{fig:Fig7} . We find a crossover between critical and non-critical behavior when we change the value of the aspect ratio. The behavior of the frequency distribution of avalanches for large values of this parameter - elongated grains - is consistent with self-organized criticality, while smaller values of the aspect ratio lead to non-critical behavior. Results are shown for simulations with the values for the aspect ratio $(q = 1.04)$ and $(q = 1.4)$ respectively. This result is in qualitative agreement with the real-life experiment of Ref. \cite{Frettea}.

\section{Conclusions}

We have investigated the behaviour of earthquake models which are known to produce avalanches and found that the phenomenological Gutenberg-Richter scaling law as well as the power law found in models of the Burridge-Knopoff type can be easily reproduced. An advantage of our ``train'' model is that it is relatively simple, depending only on a static friction which varies randomly from point to point over the contact surface of the fault we wish to model, with no dependence on velocity. Our models are easily simulated on any computer, with one feature being the ease with which the parameters can be changed and different regimes investigated.

These kinds of models may be useful for the identification of the details of the physics of earthquake events which are important for the understanding of the behaviour of the global statistics, in an area which is difficult to investigate experimentally, in particular by helping to identify which are the important features of the real events which need to be included.

The stochastic train model may be mapped onto a sandpile model where the critical slope is drawn from some statistical distribution, allowing us to control the features of this distribution and to study e.g. the influence of its width on the behaviour of the system. Finally, by allowing the local slope to be real-valued, we were able to investigate the effect of the aspect ratio of the grains on the statistical distribution of the avalanches and found that these distributions are consistent with critical behaviour only for elongated grains, as was shown in experiments with ricepiles. A more detailed analysis of the real-valued stochastic sandpile will be published elsewhere.

\section*{Acknowledgements}

We acknowledge financial support from the Brazilian agencies CNPq and
FAPERJ.  We would like to thank M.~K.~Olsen,
M.~A.~F.~Menezes and T.~J.~P.~Penna for discussions and suggestions
which helped to improve the manuscript.

\end{document}